\documentstyle[preprint,revtex]{aps}
\begin{document}
\draft
\preprint{SCCS-527}

\begin{title}
On generalized cluster algorithms for frustrated spin models
\end{title}
\author{P.D.~Coddington}
\begin{instit}
Northeast Parallel Architectures Center, 111 College Place, \\
Syracuse University, Syracuse, NY 13244, U.S.A.
\end{instit}
\author{L.~Han}
\begin{instit}
Physics Department, Syracuse University, Syracuse, NY 13244, U.S.A.
\end{instit}
\begin{center}
\bigskip
November 5, 1993
\bigskip
\end{center}
\begin{abstract}
Standard Monte Carlo cluster algorithms have proven to be very effective for
many different spin models, however they fail for frustrated spin systems.
Recently a generalized cluster algorithm was introduced that works
extremely well for the fully frustrated Ising model on a square lattice,
by placing bonds between sites based on information from plaquettes rather
than links of the lattice.
Here we study some properties of this algorithm and some variants of it.
We introduce a practical methodology for constructing a generalized
cluster algorithm for a given spin model, and investigate apply this
method to some other frustrated Ising models.
We find that such algorithms work well for simple fully frustrated
Ising models in two dimensions, but appear to work poorly or not at all
for more complex models such as spin glasses.
\end{abstract}
% \pacs{PACS numbers 05.50.+q, 75.40.Mg}

\vfill

\newpage
\narrowtext

\section{INTRODUCTION}
\label{sec:intro}

The cluster update algorithms of Swendsen and Wang~\cite{SW} and
Wolff~\cite{wolff} can provide a great improvement in computational efficiency
over the Metropolis algorithm~\cite{metrop,monte} and other local Monte Carlo
update schemes.
Due to the non-local nature of the cluster algorithms, which update large
clusters of spins at a time, they are very effective at decorrelating
successive configurations and can therefore greatly
reduce critical slowing down~\cite{monte,sokalrev}
in certain ferromagnetic and anti-ferromagnetic spin models.
However the cluster algorithms are ineffective for frustrated spin
models~\cite{frustrated} such as spin glasses~\cite{glass},
the systems for which critical slowing down can be most extreme,
and for which a non-local update algorithm would be of most benefit.
The reason for this failure is that at the
critical point, these algorithms produce a cluster
that encompasses a large fraction (if not all) of the sites in
the lattice, so that updating the spins in this cluster is virtually the
same as a trivial global spin update.

Kandel, Ben-Av and Domany (KBD) introduced a generalized cluster
algorithm~\cite{kandel90,kandel91} that includes the other cluster
algorithms as special cases. Their method also works very well for the two
dimensional fully frustrated Ising model~\cite{kandel90,kandel92},
but has not been applied to any other frustrated spin models.
The only other cluster algorithm that has been shown to work effectively for
a frustrated spin model is the Replica Monte Carlo algorithm~\cite{replica},
however this has only been successfully applied to the 2-$d$ Ising spin
glass~\cite{repglass}.
It is therefore of great interest to determine whether the generalized
cluster algorithm can be applied to other frustrated spin models.
We have investigated some variants of the KBD algorithm for the 2-$d$
fully frustrated Ising model (FFIM), and attempted to apply the method
to some other frustrated two dimensional Ising models.

We introduce a practical method for constructing a generalized cluster
algorithm for a given spin model in section~\ref{sec:algorithm}.
We then apply this methodology to
the FFIM on a square lattice in section~\ref{sec:square},
the triangular lattice Ising anti-ferromagnet in
section~\ref{sec:triangle}, and
the 2-$d$ Ising spin glass in section~\ref{sec:others}.
We analyze the performance of the generalized cluster algorithms for
each of these models.

\section{THE GENERALIZED CLUSTER ALGORITHM}
\label{sec:algorithm}

First we briefly describe the Swendsen-Wang (SW) cluster algorithm for the
Ising model~\cite{monte}, which has an interaction energy given by
\begin{equation}
E = - \sum_{<i,j>} J_{ij} \sigma_{i} \sigma_{j},            \label{eq:energy}
\end{equation}
where the spins $\sigma_{i}$ can take the values $+1$ or $-1$. We will consider
the case where the interaction strength $J_{ij}$ takes on values $+J$ or $-J$,
for some constant parameter $J > 0$.
Let us define a link to be the connection between two neighboring sites on
the lattice. A link is said to be satisfied if it is in a state of minimum
energy, which means that the spins on the two sites are the same for
$J_{ij} > 0$ or opposite for $J_{ij} < 0$, otherwise it is unsatisfied.
In the SW algorithm, bonds are introduced between spins on neighboring sites
with probability $1 \! - \! e^{-2K}$ if the link is satisfied, and 0 if it
is unsatisfied. Here $K = J/kT$ is a dimensionless coupling constant, where
$T$ is the temperature and $k$ is Boltzmann's constant.
This procedure creates clusters of bonded sites. The update part of the
algorithm consists of flipping all the spins in a cluster
(i.e. $\sigma_{i} \rightarrow -\sigma_{i}$)
with probability $1 \over 2$.
Since the clusters can be quite large, this clearly produces large, non-local
changes in the spin configuration, and thus decorrelates the configurations
much faster than local update algorithms.

This algorithm works very well for the ferromagnetic ($J_{ij} = +J$)
and anti-ferromagnetic ($J_{ij} = -J$) Ising model, but fails for frustrated
systems, which have a mixture of positive and negative couplings in such
a way that it is not possible for all links to be satisfied in the ground
state. The simple reason for this failure
is that the critical point for such systems usually occurs at low temperature
(i.e. large $K$), where the probability of putting a bond on a satisfied
link is close to 1. For example, the two dimensional fully frustrated Ising
model has a critical point at zero temperature, where there are 3 satisfied
links per plaquette. Hence the SW algorithm bonds each site to 3 of its
neighbors on average, which results in all the sites in the lattice being
connected into a single cluster.  A more fundamental reason for
this problem is that the SW algorithm only uses information from
links, from which it is not possible to see the frustration in the system.
To do that one needs information from the spins over at least a plaquette,
which is the basic idea behind the KBD algorithm.

The generalized cluster algorithm begins by expressing the energy of the
system as a sum $E = \sum_{l} V_{l}$.
We will consider $l$ to label subregions of the lattice such as links,
plaquettes, or some other group of sites that can give a tiling of the
lattice, that is, they can be replicated over the lattice in such a way
that each link is part of one (and only one) subregion.
We will refer to such a subregion as a {\it tile}.
For each tile $l$ we stochastically assign one of $n$ possible operations with
a probability $P_{i}^{l}(u)$ that depends on the spin configuration $u$
and the operation $i$. These operations involve placing bonds between
certain sites in the tile $l$.
We refer to the placement (or non-placement) of bonds as {\it freezing}
(or {\it deleting}) the connections between two neighboring sites.
The frozen bonds produce clusters of connected sites.
The probabilities must of course satisfy
\begin{equation}
\sum_{i} P_{i}^{l}(u) = 1, \qquad 0 \le P_{i}^{l}(u) \le 1,  \label{eq:norm}
\end{equation}
for all $l$ and $u$.
KBD show that detailed balance is satisfied if the probabilities
also satisfy~\cite{kandel91}
\begin{equation}
E(u) - {1 \over \beta} \log P_{i}(u) + C_i =
\left\{ \begin{array}{ll}
	0      \quad & \mbox{for an allowed operation $i$} \\
        \infty \quad & \mbox{otherwise.}
	\end{array}
\right.                                                       \label{eq:detbal}
\end{equation}
Here $E(u)$ is the energy for the configuration $u$,
$C_i$ is a constant (independent of the configuration),
$\beta = 1 / kT$ is the inverse temperature, and we have
suppressed the subscript $l$ labeling the particular tile.
To simplify matters we will concentrate on a special case of the KBD
generalized cluster algorithm, and define an operation as {\it allowed}
if it does not freeze any unsatisfied links of the configuration $u$
(for a more general definition, see Ref.~\cite{kandel91}).
If we define $a_i(u)$ to be 1 if operation $i$ is allowed for configuration
$u$, and 0 otherwise,
then equation~\ref{eq:detbal} can be written as
\begin{equation}
P_{i}(u) = a_{i}(u) \, e^{\beta(E(u) + C_i)}.               \label{eq:probs}
\end{equation}
Imposing the normalization condition of equation~\ref{eq:norm}, we have
\begin{equation}
\sum_{i} a_i(u) \, c_i = e^{-\beta E(u)}  \qquad \forall \, u\, ,
\qquad \quad c_i = e^{\beta C_i}.                           \label{eq:normprob}
\end{equation}
This can be conveniently written as a matrix equation
\begin{equation}
{\bf A} \, {\bf c} = {\bf b},           \label{eq:matrix}
\end{equation}
where the $c_i$ are the elements of the vector {\bf c},
the Boltzmann factors $b_j = e^{-\beta E(u)}$ are the elements of the vector
{\bf b}, and the $a_i(u)$ make up the elements $A_{ij}$ of the
{\it allowance matrix} {\bf A},
with $j$ labeling the different possible configurations $u$.

Formulating a generalized cluster algorithm involves identifying some
candidate operations, constructing the allowance matrix {\bf A},
and then solving the
matrix equation~\ref{eq:matrix} for {\bf c}, from which the
probabilities for each operation can be calculated via equation~\ref{eq:probs}.
For a given set of operations and configurations, there is no guarantee that
a solution to the matrix equation will exist.
Even if the equation has a solution, it is quite likely that it will not
satisfy the constraint that all the probabilities must be between 0 and 1.
For all but the simplest models and tiles, the number of possible
operations will be large, and the difficulty in this method is in choosing
some subset of operations that allow a valid solution to the matrix
equation. Here we offer some guidelines for choosing operators in order to
maximize the possibility of obtaining a valid algorithm. Examples of this
procedure will be given in the following sections.

Clearly the simplest way to approach the problem is to choose the operations
so that the allowance matrix is square (the same number of operations as
configurations) and upper triangular.
This will guarantee a solution, although it may not satisfy the constraint
that the probabilities are all between 0 and 1.
A quick preliminary check to see if this constraint is violated is to check
whether any of the elements of the solution vector {\bf c} are
negative, which is not allowed since $c_i = e^{C_i}$.

There will generally be more than one operation available for a given
configuration which will produce an upper triangular allowance matrix.
For frustrated
models, we want to choose the operation which freezes the least number of
bonds, and bonds together the fewest sites into the same cluster. However
in some cases it is advantageous to choose more than one such operation for
a given configuration. This gives a system of equations that is
under-determined, that is, there are more unknowns than equations, producing
free parameters in the solution. Having free parameters to play with is
extremely useful: firstly, we may change an invalid solution into one that
obeys the probability constraints; secondly, it may allow some
optimization in the algorithm by tuning the size of the largest
cluster, as done by KBD for the FFIM~\cite{kandel90,kandel92}.
Also, if an operation $i$ only occurs for a single configuration, there are
no constraints on the choice of $C_i$, which also gives a free parameter
in choosing the probabilities.

For frustrated spin models, we are usually interested in simulating the
system at low temperature. A good method of constructing an algorithm in
this case is to start with a simpler zero temperature algorithm. In this
case we only need consider the ground state configurations, and those
operations that conserve the energy of these configurations. This greatly
simplifies the problem of determining valid solutions, and also allows a
quicker indication of whether the algorithm can produce a good distribution
of cluster sizes. If the implemented zero temperature algorithm produces
clusters that are neither too large nor too small, then an extended
algorithm should work even better at non-zero temperature where the
largest cluster will be smaller.

It should be possible to automate the procedure for constructing a valid
generalized cluster algorithm.  Given a particular spin model,
a particular lattice, and a particular tiling of the lattice,
a program could generate all possible configurations of the tile,
and all possible allowed operations on those configurations. The program could
then pick one possible choice of operations that give an upper triangular
allowance matrix, possibly including some redundant operations to allow free
parameters in the solution. It could then solve the matrix equation, calculate
the probabilities, and check that the constraints on them are satisfied
(this could all be done using a symbolic algebra package such as
Mathematica~\cite{mathematica}).
If not, it could choose a new set of operations and repeat the procedure,
until it finds a valid solution, or exhausts the list of operations.
So far we have done most of this process by hand, however the steps involved
appear amenable to automation.

\section{THE SQUARE LATTICE FULLY FRUSTRATED ISING MODEL}
\label{sec:square}

The fully frustrated Ising model (FFIM) was introduced by
Villain~\cite{villain},
as a simple regular frustrated system lacking the extra complication
of disorder that is present in systems such as spin glasses.
On a square lattice, the FFIM has ferromagnetic couplings on all links
except for a line of anti-ferromagnetic couplings on every second column
of links. This means that every plaquette is frustrated, having 3
ferromagnetic links and one anti-ferromagnetic link.

The critical point of this model is at zero temperature, where the
configurations have one unsatisfied link per plaquette.
The SW algorithm for this model puts a bond on all satisfied links,
so there is only one unbonded link per plaquette. It is easy to prove
(using the method of Kandel and Domany~\cite{kandel92}) that in this
case all the sites are bonded into a single cluster.
Consider the dual lattice, where there is a site centered in each plaquette,
and dual links which connect these dual sites, so that each dual link crosses
(at right angles) a single link of the original lattice. Let us put bonds on
the dual links which cross through unbonded original links,
and no bonds on dual links which cross bonded original links.
Now suppose that in the original lattice there was a site or cluster
of sites that was not bonded to the main cluster. Then on the dual lattice,
it would be surrounded by a closed loop of dual bonds. However it is easy
to see that for this model it is impossible to construct such a closed loop.
Since there is only one unbonded link per plaquette on the original lattice,
there can only be one dual bond coming from any dual site at the center of a
plaquette. It is therefore impossible to connect more than one dual bond,
since that would require two dual bonds coming from the same dual site.
Hence dual sites are only connected in pairs, by a single dual
bond, and there is no possibility of making a closed loop of dual bonds.

Thus the SW algorithm bonds all the sites in the lattice together into a
single cluster at the critical point, and so does not work for this model.
However the generalized cluster algorithm of KBD, which deletes bonds using
information on the state of plaquettes rather than links, works extremely
well in this case~\cite{kandel90,kandel92}.

\subsection{The KBD Algorithm}
\label{subsec:kbd}

We will derive the KBD algorithm in a slightly different way, so as to
make the generalization to other frustrated models more obvious, and
to highlight the simple matrix methodology outlined in
section~\ref{sec:algorithm}.

In the KBD algorithm, the tile used is a plaquette.
There are seven freeze/delete operations, as
shown in Fig.~\ref{ops_sq}.
Operations 2-7 conserve the energy of a configuration.
Note that operations 2-3 and operations 4-7 are just different orientations
of the same basic topology of frozen and deleted bonds.
To simplify matters, we will classify the operations into the 3 basic
topologies ({\tt A}, {\tt B}, and {\tt C} in Fig.~\ref{ops_sq}) which we
will refer to as {\it operators}. Each of the operators has some number
of possible orientations: 1 for operator {\tt A}, 2 for operator {\tt B},
and 4 for operator {\tt C}.
However, the important feature of these operators for constructing a
generalized cluster algorithm is the number of orientations that are possible
for any given configuration of spins in a plaquette.
For example, consider a plaquette having the lowest energy, $-2J$. This has one
unsatisfied link. Since we can only freeze satisfied links, this means there
is only one allowed orientation of operator {\tt B} (the
one with the deleted bonds corresponding to the unsatisfied link and the
link parallel to it), and two allowed orientations of operator {\tt C}
(deleting the unsatisfied link and one of the links at right angles to it
on either side).

Likewise, the important property of a plaquette configuration is not the spins,
but rather which of the links are satisfied, since that alone determines which
freeze/delete operations can be applied. From now on we will use the term
``configuration'' to mean a
configuration of satisfied and unsatisfied links, rather than spins.
Again, there are only a few basic topologies of
satisfied and unsatisfied links -- for the square lattice FFIM there are
only two (either 1 or 3 unsatisfied links), in general there will be at
least as many as there are possible energy states of the tile.
All other configurations are just different orientations of these basic
``link topologies''.

To simplify matters, we will consider different orientations of the
configurations and the freeze/delete operations to be equivalent.
In order to use the matrix formalism outlined in section~\ref{sec:algorithm},
we need to know the number of different link topologies, and the
number of allowed orientations $n_{ij}$ of each operator $i$ for any given
configuration of link topology $j$. If we adopt this approach, the
elements of the allowance matrix {\bf A} will be $n_{ij}$.
The allowance matrix for the KBD algorithm is shown in
Table~\ref{tab:matrix_sq}.

Solving equation~\ref{eq:matrix} (which is easily done using Mathematica)
gives the probabilities $P_{i}(j)$ given in Table~\ref{tab:probs_sq}.
Here the $P_i(j)$
refer to the probability of choosing operator $i$ for a given
configuration of link topology $j$.
Each orientation of the operator which is allowed for the particular
configuration would then be assigned with equal probability, $1 / n_{ij}$.
For example, if we chose operator {\tt C}, we would randomly
choose with probability $1 \over 2$ to delete either of the 2 bonds
perpendicular to the unsatisfied link.

Since for this model we have 3 operators and only 2 link topologies,
the solution to the matrix equation has a free parameter $p$.
KBD use this freedom in the choice of the probabilities to prove
(using a subtle geometrical argument) that for a particular choice of
probabilities, the lattice is always split into at least two large clusters
at the zero temperature critical point~\cite{kandel92}.
For the probabilities chosen by KBD, the average maximum cluster size is
neither too large nor too small (the average being 0.6432(5) for a
$64^2$ lattice), so the KBD cluster algorithm works extremely well.

When we set up the matrix equation, we could have treated all possible
orientations of the configurations and the operators as different, rather
than grouping them together as we have done. This would give 4
lowest energy configurations (rather than just 1), corresponding to the
4 possible positions of the single unsatisfied link in the plaquette.
It would also give the 6 energy conserving operations of
Fig.~\ref{ops_sq} (rather than the 2 basic operators).
So instead of a single free parameter, from having 2 operators
and only 1 configuration, there are now 2 free parameters, from having
6 operations and only 4 configurations. If we solve the extended
matrix equation, we find that the free parameter enters (as one might expect)
in the probability of choosing which of the bonds perpendicular to the
unsatisfied link are to be deleted for operator {\tt C}.
We now have an extra probability $p_{sub}$ of choosing the operations from
the subgroup (4-5) of the angled operations,
and $1-p_{sub}$ of choosing the operations from the subgroup (6-7),
where previously we had $p_{sub} = {1 \over 2}$.

Both these approaches -- using the basic topologies or using all possible
orientations --  are correct, i.e. satisfy detailed balance.
The first is simpler and allows an easier implementation
of more complex problems where there are a large number of possible
topologies and orientations;
the other may provide some extra freedom in the choice of probabilities,
that can allow some tuning to improve the performance of the algorithm,
as we shall see in section~\ref{subsec:var_kbd}.

Although the KBD algorithm satisfies detailed balance, it is not ergodic
at zero temperature -- that is, starting from a particular ground state
configuration, it cannot generate all other ground state configurations.
In fact, it can readily be seen for small lattice sizes that the KBD algorithm
cycles between subsets of possible ground state configurations. In order for
the algorithm to be ergodic, each KBD update must be followed by a Metropolis
sweep.
However, to our knowledge ergodicity of the Metropolis algorithm has not been
{\it proven} for this model at zero temperature, although it is believed
to be true (proving ergodicity is often a very difficult problem, see for
example Ref.~\cite{ergodic}).
It can however be easily seen (again by just looking at test
configurations on small lattices) that the Metropolis algorithm is {\it not}
ergodic if the sites to be updated are chosen in serial (rather than random)
order.

\subsection{Dynamic Critical Exponents}
\label{subsec:dynamics}

KBD measured the dynamic critical exponent for the exponential autocorrelation
time to be $z \approx 0.55$ for their algorithm~\cite{kandel92},
compared to $z \approx 2$ for the Metropolis algorithm
(see Refs.~\cite{monte,sokalrev} for discussion and definitions of
autocorrelation times and dynamic critical exponents).
However they had very low statistics (100,000 sweeps) and fairly small
lattice sizes (up to $128^2$), and it has been seen for the Ising ferromagnet
that it is very difficult to get an accurate determination of $z$ from
such data~\cite{paul}.
We have therefore obtained much better data for the autocorrelations for
this algorithm, in order to get a better determination of the dynamic
critical exponent.

We measured the normalized autocorrelation function
\begin{equation}
\rho_A(t) = {<A(0)A(t)> - <A(0)>^2 \over <A(0)A(0)> - <A(0)>^2}
\label{eq:rho}
\end{equation}
for the magnetization ($A = M$) as well as
the square and the absolute value of the magnetization
($A = M^2$ and $A = |M|$),
and used this to extract exponential autocorrelation times $\tau_{exp}$, via
\begin{equation}
\rho(t) \sim e^{-t/\tau_{exp}},
\label{eq:tau_exp}
\end{equation}
as well as integrated autocorrelation times
\begin{equation}
\tau_{int} = {1 \over 2} + \sum_{t=1}^{\infty} \rho_A(t),
\label{eq:tau_int}
\end{equation}
that are relevant to the increase in the statistical error due to
correlated configurations.
Details of the methods we used for measuring the autocorrelations, doing the
fits, and estimating the errors are given in Ref.~\cite{paul}.

In Table~\ref{tab:tau_kbd} we give the results for the autocorrelation times
of $M^2$ for different lattice sizes $L$ up to $L=256$.
All the results are from at least $7.5 \! \times \! 10^6$ iterations
(more for smaller lattice sizes).
Fig.~\ref{tau_kbd} shows a log-log plot of these results.
Straight line $\chi^2$ fits give
$z_{int} = 0.28(1)$ and
$z_{exp} = 0.66(5)$.
The integrated autocorrelations also fit fairly well to a logarithm,
so that $z_{int}$ could also be zero.
This uncertainty also occurs with the autocorrelations of cluster algorithms
for the ferromagnetic Ising model~\cite{paul}, since it is very
difficult to differentiate between a logarithm and a small power.

Let us define $\tau_{0}$ to be the autocorrelation time obtained from
an exponential fit to $\rho(t)$ at only two points, $t=0$ and $t=1$.
This is also plotted in Fig.~\ref{tau_kbd}.
We can see that $\tau_{0}$ grows extremely slowly with $L$ (slower than
$\log(\log L)$ in fact), and seems to be approaching a constant value.
This is in marked contrast to $\tau_{exp}$, which grows as a
substantial power of $L$. Unfortunately this makes it very
difficult to get a good asymptotic fit to $\rho(t)$ at large $t$, since it
has already fallen off so much at $t=1$ that the signal is very small and
noisy in the region where we need to fit to obtain $\tau_{exp}$. The fact
that $\tau_{0}$ increases so slowly means that the $\tau_{int}$ grows
much slower than $\tau_{exp}$, and consequently $z_{int}$ is much
smaller than $z_{exp}$.

The reason that $\tau_{int} << \tau_{exp}$ is presumably that
the operator $M^2$ does not have a large overlap with the slowest mode.
The value of $\tau_{exp}$ should be independent of the operator measured,
since it measures the relaxation of the slowest mode,
however if there is not good overlap with this mode, this
may mean that our estimate of $\tau_{exp}$ is not a good one.
To get a better understanding of the dynamics of this system, one would
need to find an operator for which $\tau_{int} \approx \tau_{exp}$.
We also measured the autocorrelation times for the magnetization and
the absolute value of the magnetization. For the former the results
were zero for all lattice sizes, while the latter gave results almost
identical to those given above for the square of the magnetization.

In almost every iteration of the standard KBD algorithm, the lattice is
split into 2 clusters, and only rarely into 3 or more. The update consists
of flipping the spins in a cluster with probability 1/2. If there are two
clusters, this means that in half of the iterations either no spins are
flipped, or all the spins are flipped, so there is no change in
the configuration apart from a trivial global spin flip. Only half of the
iterations make a non-trivial update of the configuration by
updating one of the clusters and not the other. If there are 2 clusters,
a more effective update scheme is to pick one of them at random and flip
its spins. This guarantees that there are no ``wasted'' iterations.
One would expect that this new update scheme would decrease the error
in measurable quantities by a factor of $\sqrt 2$, due to a decrease in
$\tau_{int}$ by a factor of 2.
However it turns out that the improvement is actually greater than this.
The reason is that the the autocorrelation function for the new algorithm
{\it alternates in sign}, as shown in Fig.~\ref{new_kbd}.
Thus using the definition for $\tau_{int}$ (equation~\ref{eq:tau_int})
gives a value close to zero,
and much less than half of the value for the standard KBD algorithm
with an exponential autocorrelation function.

The autocorrelation function for a Markov process can generally be expressed
asymptotically as ${\rho(t) = a^t}$, where ${|a| < 1}$~\cite{timeseries}.
If ${a > 0}$, then redefining ${a = e^{-1/\tau_{exp}}}$ gives the standard
asymptotic exponential form for $\rho(t)$.
However it is also possible to have $a < 0$,
in which case the autocorrelation function still falls off exponentially,
but its sign is $(-1)^t$. This generally indicates that the original data
forms an alternating time series, with successive measurements falling
predominantly on one side and then the other of the mean
value~\cite{timeseries}.
If ${\rho(t) = a^t}$, then $\tau_{int} = {1 \over 2}{{1+a} \over {1-a}}$,
which can be very small for $a$ negative.
In order to calculate $\tau_{int}$ for the alternating autocorrelation function
from the simulation data, we adopt a similar procedure to that used for
the standard autocorrelation function~\cite{paul}.
We treat $\rho(2t)$ and $-\rho(2t+1)$ as separate positive functions,
and fit them in the usual way to find $\tau_{exp}^{+}$ and $\tau_{exp}^{-}$.
These two values are very similar for this algorithm, as are the corresponding
dynamic critical exponents
$z_{exp}^{+} = 0.42(4)$ and $z_{exp}^{-} = 0.36(4)$, which are distinctly
smaller than $z_{exp}$ for the standard KBD algorithm.
A logarithmic increase in $\tau_{exp}$ with lattice size
(i.e. $z_{exp}^{\pm} = 0$) is also consistent with the data.

We can also calculate
\begin{eqnarray}
\tau_{int}^{+} & = & {1 \over 2} + \sum_{t=1}^{\infty} \rho(2t), \nonumber \\
\tau_{int}^{-} & = & - \sum_{t=0}^{\infty} \rho(2t+1),
\label{eq:tauneg1}
\end{eqnarray}
in the usual way by
by splitting each infinite sum into a small $t$ finite sum plus the remaining
large $t$ infinite sum, which can be summed analytically using the fact that
the autocorrelation functions fall off asymptotically
as $A e^{-t/\tau_{exp}}$.
This gives
\begin{eqnarray}
\tau_{int}^{+} & = & {1 \over 2} + \sum_{t=1}^{W^{+}} \rho(2t) +
                   {A^{+} e^{-2(W^{+}+1)/\tau_{exp}^{+}} \over
                   1 - e^{-2/\tau_{exp}^{+}}}, \nonumber \\
\tau_{int}^{-} & = & \sum_{t=0}^{W^{-}} \rho(2t+1) +
                   {A^{-} e^{-[2(W^{-}+1)+1]/\tau_{exp}^{-}} \over
		   1 - e^{-2/\tau_{exp}^{-}}},
\label{eq:tauneg2}
\end{eqnarray}
where the windows $W^+$ and $W^-$ are taken to be the end of the region where
we fit for $\tau_{exp}$. Now we can obtain the integrated autocorrelation time
for the alternating $\rho(t)$ as
\begin{equation}
\tau_{int} = \tau_{int}^{+} - \tau_{int}^{-}.
\label{eq:tauneg3}
\end{equation}
The integrated autocorrelation times for this method are shown in
Fig.~\ref{tau_kbd}. They are smaller than for the standard KBD algorithm,
however the dynamic critical exponent $z_{int} = 0.32(2)$
is approximately the same, and could also be zero.

Another interesting result concerns the autocorrelation function $\rho(t)$ for
the KBD algorithm {\it without} the Metropolis sweep needed for ergodicity,
which is shown in Fig.~\ref{rho_ergodic}, along with $\rho(t)$ for the
algorithm {\it including} Metropolis sweeps. The latter falls off
asymptotically as an exponential, as expected,
while $\rho(t)$ for the non-ergodic algorithm asymptotes to a logarithm.
This implies that the autocorrelation time is infinite, which one might
expect for a non-ergodic algorithm (although it is not clear why $\rho(t)$
should have a perfectly logarithmic form).
However, measuring the autocorrelation function at zero temperature cannot
tell us whether an algorithm is ergodic.
For example, the Metropolis algorithm implemented so that the sites are
visited in serial order is also a non-ergodic algorithm for this model,
although we found that its autocorrelation function was asymptotically
exponential, and in fact has a much {\it smaller} apparent autocorrelation
time than the algorithm with the sites chosen in random order, which is
presumed to be ergodic.
In this case the non-ergodic algorithm has smaller correlations between
successive configurations, but only generates a subset of all possible
configurations.

% \newpage

\subsection{A Variant of the KBD Algorithm}
\label{subsec:var_kbd}

In order to guarantee that the lattice is broken up into at least 2 large
clusters at zero temperature, KBD used the freedom in choosing the
probabilities so that only operations 2 and 3 of Fig.~\ref{ops_sq}
had non-zero probability at ${T \! = \!0}$.
However in order to investigate the effect of tuning the extra free
parameter $p_{sub}$, we used an alternate version of the KBD algorithm
that uses only the angle operations (4-7) at zero temperature,
rather than the operations (2,3) used by KBD.
We measured the cluster sizes and autocorrelations for this algorithm
for different values of $p_{sub}$: 1.0, 0.8, and 0.5. As noted by KBD, for
this choice of probabilities the lattice is no longer split into two
large clusters, rather we get one very large cluster and a number of very
small clusters.
Tuning this extra free parameter does have an effect
on the largest cluster size, although it is quite small:
the average maximum cluster size is
0.91762(1) of the lattice for $p_{sub} = 0.5$,
0.89978(6)  for $p_{sub} = 0.8$, and
0.87371(10) for $p_{sub} = 1.0$.
For $p_{sub} = 1.0$ the geometry of the operators is such that all sites
are either in a single site cluster or part of a spanning cluster
(the largest cluster).

Integrated autocorrelation times for these three variations of the KBD
algorithm are shown in Fig.~\ref{tau_all},
along with results for the Metropolis algorithm,
and the KBD results from Fig~\ref{tau_kbd}.
The results are quite surprising. As expected, $p_{sub} = 0.5$ gives a value
$z = 1.97(2)$ which is very similar to the Metropolis algorithm,
since the largest cluster encompasses most of the lattice and consequently
the cluster update does little apart from a global spin flip.
For larger values of $p_{sub}$ we have seen that
the biggest cluster size is reduced very slightly, however this
is enough to substantially reduce the autocorrelations:
$p_{sub} = 0.8$ has $z = 0.62(2)$, while $p_{sub} = 1.0$ has $z = 0.48(1)$.
It is very surprising that in going from $p_{sub} = 0.5$ to 1.0 the biggest
cluster size is only reduced by about $5\%$, yet $z$ is reduced from
around 2.0 to around 0.5!
This result is very promising, since it implies that critical slowing down
can be substantially reduced even when the biggest cluster size is quite
large ($87\%$ of the lattice in this case), so for other frustrated spin
models we may not have to unfreeze the lattice very much to get good results.

Our result is also surprising in that $z$ appears to vary continuously
between the Metropolis and KBD values. Based on the ideas of dynamic
universality, it is more likely that there is a critical value of $p_{sub}$
for which $z$ jumps from one value (Metropolis) to the other (KBD).
We have used rather small lattices ($L \le 64$) for our analysis,
so it is quite possible that the true value of $z$ is actually the same for
$p_{sub} = 0.8$ and $p_{sub} = 1.0$.

The generalized cluster algorithm works extremely well for the square lattice
FFI, almost completely eliminating critical slowing down. However it is not
clear that it can be successfully applied to other frustrated models. It is
only by a fortuitous geometrical happenstance that the standard KBD
algorithm is able to split the lattice into two large clusters at the critical
point, and the algorithm we have investigated which uses only the angled
operations works well only because
it has a tunable parameter $p_{sub}$ that allows the maximum cluster size
to be reduced just enough to give a greatly reduced $z$. We were therefore
interested to see whether the generalized cluster algorithms would work for
other frustrated spin models.

\section{THE TRIANGULAR LATTICE ISING ANTIFERROMAGNET}
\label{sec:triangle}

We first attempted to apply the KBD cluster algorithm to an even simpler
fully frustrated model, the anti-ferromagnetic Ising model on a triangular
lattice. This model is very similar to the square lattice FFIM,
being in the same universality class and also having a
critical point at zero temperature~\cite{wannier,stephenson}.

The ground state of this model has only one unsatisfied link per triangular
plaquette. The SW algorithm puts a bond on all satisfied links, so we can
apply the same argument as for the square lattice FFIM to see that the SW
algorithm again freezes the lattice into a single cluster at zero temperature.

We can get some idea of the percentage of bonds that need to be deleted
to break up this single cluster by noting that the bond percolation threshold
for a triangular lattice is 0.35792~\cite{perc}.
For the square lattice the bond percolation threshold is 0.5~\cite{perc},
which is also the ratio of frozen bonds to links in the KBD algorithm for
the square lattice FFIM at the zero temperature critical point.
This is of course only a rough pointer to what is required, since for bond
percolation the bonds are placed on the lattice at random, whereas for the
KBD algorithm on the square lattice we must have two bonds for each plaquette
at zero temperature.
Even with the bond/link ratio being the same as the percolation threshold,
the largest cluster for the KBD algorithm percolates at a temperature well
above the critical point.

\subsection{The Plaquette Algorithm}
\label{subsec:triplaq}

The application of the KBD algorithm to the triangular lattice FFIM is
very simple. As with the square lattice, we choose the basic element to be a
(triangular) plaquette, which has two possible energy states, $+J$ and $-J$.
For the triangular lattice there are only four possible freeze/delete
operations, which are shown in Fig.~\ref{ops_tri}.
We can either delete all the bonds, or delete one of the two satisfied bonds.
Since the ground state configuration has a single unsatisfied link, and
the energy conserving operations (2-4 of Fig.~\ref{ops_tri}) freeze a
single bond, there are 3 possible orientations of each. Thus, unlike the
square lattice case, there are no free parameters when we solve the matrix
equation~\ref{eq:matrix}, even when all possible orientations
of the configurations and operations are used.
The probabilities for each operation are given in Table~\ref{tab:probs_tri}.

At the zero temperature critical point, we only perform the energy conserving
operations (2-4). This gives a bond/link ratio of $1 \over 3$,
which is below the percolation threshold for the triangular lattice,
although in this case the bonds are not placed randomly,
but rather one per plaquette.
However we expect to at least be able to create multiple clusters,
as with the KBD algorithm with angled operations for the square lattice FFIM.

We tested this algorithm at zero temperature, following every cluster update
by a Metropolis sweep to ensure ergodicity.
We found that the algorithm produces one very large cluster,
and a number of very small clusters.
For a $64^2$ lattice there are an average of 205.1(1) clusters, with the
average size of the largest cluster being 0.9233(1) of the lattice volume.
This is very similar to the results for the square lattice case using the
angled operations with equal probability. In that case there is a free
parameter ($p_{sub}$) that allows us to bias the choice of the operations so
as to reduce the largest cluster size and greatly improve the performance of
the algorithm. However for the triangular lattice there is no such freedom,
and in order
to satisfy detailed balance the two possible energy conserving operations
for each plaquette must be chosen with equal probability. We checked that
introducing a bias in the choice of the operations (2-4) does indeed reduce
the largest cluster size,
but of course it also gives incorrect results due to the violation of
detailed balance.

Since the ground state for this model is paramagnetic, we measured the
paramagnetic susceptibility ${\chi = V <M^2>}$,
where $M$ is the average magnetization per site,
and $V$ is the lattice volume. At the zero temperature critical point, this
quantity approaches a constant value as the lattice size is increased.
We also measured the spin correlation function
${\Gamma(R) = <\sigma_0 \sigma_R>}$ (which is known exactly for an infinite
lattice~\cite{correlations}) at a distance $R\!=\!2$.
We measured the autocorrelations in $\chi$ and $\Gamma(2)$
for both the plaquette algorithm and the standard Metropolis algorithm.
The autocorrelation times for the plaquette algorithm were doubled to
give a fairer comparison with the Metropolis algorithm, since every iteration
of the plaquette algorithm includes a Metropolis update to ensure ergodicity.

The integrated and exponential autocorrelation times for the Metropolis
and plaquette algorithms for $\chi$ are shown in Fig.~\ref{tau_chi},
and for $\Gamma(2)$ in Fig.~\ref{tau_gamma}.
Even for the Metropolis algorithm, the autocorrelations
are quite small, and grow much slower than $L^2$.
The plaquette algorithm seems to almost completely eliminate critical
slowing down in the measurement of $\chi$, although this is not
much of an improvement over Metropolis.
However the plaquette algorithm substantially reduces the autocorrelations
for $\Gamma(2)$, especially $\tau_{exp}$, which is related to the time
required to thermalize to a ground state configuration.
Notice that $\tau_{exp}$ is different for the different operators.
Again, this means that $M^2$ (and possibly also $\Gamma(2)$) does not have
a large overlap with the slowest mode, so our measurements of this quantity
are presumably quite poor. Again, one would like to find an operator for
which $\tau_{int} \approx \tau_{exp}$.
The dynamic critical exponents for $\chi$ and $\Gamma(2)$ for the two
Monte Carlo algorithms are shown in Table~\ref{tab:z_tri}.

It is possible to construct a simple plaquette generalized cluster
algorithm for the triangular lattice FFIM that is certainly superior to the
standard Swendsen-Wang cluster algorithm, which freezes the lattice into
a single cluster and is therefore totally ineffective.
The size of the biggest cluster in the plaquette algorithm is still very large,
and there are no tunable parameters that might enable us to reduce it,
although it appears that the algorithm performs well in spite of this problem.
However this may just be an fortuitous anomaly, since even the Metropolis
algorithm performs quite well for the quantities which we measured.
We therefore tried to find a general method for improving the generalized
cluster algorithm by further decreasing the size of the largest cluster.

\newpage

\subsection{A Larger Tile Algorithm}
\label{subsec:dblplaq}

To try to improve on this algorithm, we investigated making the tile
something larger than a plaquette. Since increasing the tile from a
single link (Swendsen-Wang) to a plaquette (KBD) improves the low
temperature algorithm substantially, it is possible that an even larger
tile will produce better results.
Note that the simple double plaquette shown in Fig.~\ref{dbl_tri}(a)
cannot tile the lattice -- that is, it is not possible
to cover all the links in the lattice such that each link is uniquely
assigned to a double triangular plaquette. Since the double plaquette
has only 5 links and the number of links in the lattice is a multiple of 3,
there is an extra link required. Fig.~\ref{dbl_tri}(b) shows a set of links
that can tile a triangular lattice.

For this model a ground state configuration has two satisfied links
in each triangular plaquette.
There are 8 configurations of the tile that can occur
in the ground state, which are shown in Fig.~\ref{dblink_configs}.
Unlike the algorithm using the triangular plaquette, these ground state
configurations of the tile can have different energies.

An operation conserves energy if it freezes at least one satisfied bond
for each triangular plaquette.
For this tile, there are 17 possible energy conserving operations that
freeze 3 or fewer bonds.
These are shown in Fig.~\ref{dblink_ops}, along with the single allowed
operation that freezes 5 bonds, and one of the 5 possible operations
that freeze four bonds.
Note that operations 14-18 (and the other 4-bond operations that are not
shown) are equivalent in the sense that they ensure the same 4 sites
are bonded together into the same cluster, so there is no reason to choose
one in preference to another.
Note that there is a subtlety here in what we mean by an energy conserving
operation. These operations ensure that the total energy of a ground state
configuration is unchanged, as it should be at zero temperature. However
the energy of a {\it tile} can take three different values for a given ground
state configuration: $-4J$, $-2J$ and 0. The energy conserving operations
may change the energy of a tile, but they will not change the total energy
of the configuration (summed over all tiles).

It is quite straightforward to choose a subset of the possible operations
in Fig.~\ref{dblink_ops} which give an upper triangular allowance matrix,
as shown in Table~\ref{tab:dblink_matrix}.
Since we have many more operations than configurations, it is possible
to choose some redundant operations which add free parameters to the
probabilities.
For simplicity we have only chosen one (operation 6), however there are
other possible choices which would add extra parameters.
Solving the matrix equation gives the probabilities in
Table~\ref{tab:dblink_ops}, which give a valid algorithm at zero temperature.
Notice that if we choose to set the free parameter $p_{2}=1$,
the algorithm is greatly simplified and only involves a few operations
(1, 2, 4, 18, 19), since the probabilities for the other operations are zero.
The first three of these freeze only 2 bonds per tile, thus giving the same
bond/link ratio as the single plaquette algorithm.
The problem with this algorithm is that in order to get an upper triangular
allowance matrix (and thus sensible probabilities),
we need to choose 2 operations (18 and 19) which freeze {\it all} the bonds
in configurations 7 and 8, which make up a substantial proportion of the
tiles at zero temperature. This means that the algorithm with this larger
tile has a greater bond/link ratio than the single plaquette algorithm.
Any choice of the other operations and free parameters will suffer the same
problem.

One subtlety in this algorithm is the choice of tiling at each iteration.
For the single plaquette algorithm there are two possible tilings of the
triangular lattice, just as there are for the square lattice algorithm,
and one of the two possibilities (either the black or white squares
in a checkerboard pattern) are chosen at random for each iteration.
However in this case the tile of Fig.~\ref{dbl_tri}(b) is asymmetric,
so the number of different tilings is 12, since there are 6 possible
ways the link coming off the double plaquette can be oriented,
and 2 possible ways of partitioning the lattice
(the black and white partitioning) for each of these orientations.
One of the 12 possible tilings is chosen at random for each iteration.

We found that the larger bond/link ratio of this algorithm produced a
single cluster covering the whole lattice at zero temperature.
We tried some variations, such as taking $p_{2} < 1$, and trying different
operations, but got the same result.
Thus the generalized cluster algorithm with a larger tile does even
worse than the single plaquette algorithm for the FFIM on a triangular
lattice.

\section{OTHER 2-D FRUSTRATED MODELS}
\label{sec:others}

Let us consider the $\pm J$ Ising spin glass, where the couplings $J_{ij}$ in
equation~\ref{eq:energy} are chosen to be
ferromagnetic ($+J$) or antiferromagnetic ($-J$) at random~\cite{glass}.
For the spin glass the generalization of the KBD algorithm at the zero
temperature critical point is very simple.
For plaquettes with an even number of ferromagnetic links, we must freeze
all the satisfied links in order to conserve energy, so this is the same
as the SW algorithm.
For plaquettes with an odd number of ferromagnetic links, we have
the KBD algorithm for the FFIM.

For the spin glass there is not the regular geometry that allows the KBD
algorithm to split the lattice into two clusters for the FFIM,
and at zero temperature this combined KBD/SW algorithm
freezes the lattice into a single cluster.
It is possible that a workable generalized cluster algorithm may be
constructed for this model using a larger tile than the plaquette,
however as we have seen this was not the case for the triangular lattice FFIM.

\section{CONCLUSIONS}
\label{sec:conclusion}

One of the main ideas of KBD was to extend the basic element
of a cluster algorithm from a link connecting only two sites to a larger
geometrical object such as a plaquette.
This is clearly necessary for frustrated systems, for which knowledge of
the frustration of the system cannot be obtained from looking at the
interaction of two neighboring spins.
We have presented a simplified method for constructing a generalized
cluster algorithm for an arbitrary spin model based on any group of spins
that can tile the lattice.

However for a particular spin model, there is no guarantee that a valid
algorithm with sensible probabilities for each possible freeze/delete
operation can be constructed.
Even if such an algorithm does exist, the resulting clusters may still be
too large, causing poor performance.
For the 2-$d$ Ising spin glass we found that a generalized cluster algorithm
froze the lattice into a single cluster at zero temperature.
For the triangular lattice FFIM the method produced a very large cluster,
however the algorithm still appeared to work well.
We have also used this method to construct a generalized cluster algorithm
for the 3-$d$ cubic lattice FFIM, which also produces a very large cluster
at the critical point, and performs no better than the
Metropolis algorithm~\cite{cubic}.

Even when the plaquette algorithms do not perform well, they are still a
great improvement over the Swendsen-Wang algorithm, which is the
single-link version of the generalized cluster algorithm. It is possible
that using tiles even larger than a plaquette will improve matters further,
however we found that for the case of the triangular lattice FFIM
this produced an algorithm which froze the lattice into a single cluster
at zero temperature, and was therefore worse than the smaller tile
algorithm.

The method we have used here is a more convenient, but specialized, case
of the KBD generalized cluster algorithm. In particular,
we have looked only at the case where the bonds are either frozen or
deleted. This produces clusters that can be updated independently of
one another. However the generalized cluster algorithm allows for another
possibility, in which there is an interaction energy between the clusters.
This could allow the creation of smaller clusters that may be updated with
a probability dependent on the energies of the other clusters, in a fashion
similar to the algorithms of Niedermayer~\cite{niedermayer} and
D'Onorio de Meo {\it et al.}~\cite{demeo}.
However these kinds of interacting cluster
algorithms have not proven to be effective for non-frustrated systems,
mainly because the probability of updating a cluster is approximately
inversely proportional to the exponential of the cluster size.
Thus updating large clusters occurs with a very small probability, so
these algorithms tend to be not much better than standard local Monte Carlo
methods, at a much greater computational cost.

Although the generalized cluster algorithm seems very promising as an
approach to simulating frustrated systems,
and works very well for simple two dimensional fully frustrated Ising models,
it does not appear to be generally applicable to any spin model.
Just constructing clusters to update so as to satisfy detailed balance does
not appear to be enough -- cluster algorithms only seem to work when the
clusters are constructed in a way which reflects the physics of the model.
Thus different algorithms are required for different models, and finding
a general algorithm which works in all cases seems a daunting task.
We have not tried all possible choices of operations and parameters for
this method, however from our experience it appears that just tuning
parameters will not work, and
some new ideas are necessary before generalized cluster algorithms
can be successfully applied to
more complicated frustrated systems such as spin glasses and models in
more than two dimensions.

\acknowledgments

We are grateful to John Apostolakis, Ofer Biham, Danny Kandel, Enzo Marinari
and Alan Sokal for valuable discussions.
This work was sponsored in part by Department of Energy grants
DE-FG03-85ER25009 and DE-AC03-81ER40050,
and by a grant from the IBM corporation.

\centerline{NOTE ADDED}
\medskip
After this work was completed, we obtained the University of Marburg
preprint ``Cluster mechanisms in the fully frustrated Ising model'',
by Werner Kerler and Peter Rehberg (cond-mat/9401063), which
addresses many similar issues to the work described here.

%
% REFERENCES
%

%
% TABLES
%
\begin{table}
\caption{Matrix containing the number of possible orientations of the
freeze/delete operators of Fig.~\ref{ops_sq}
for each energy state of the the square lattice FFIM.
\vspace*{0.25truecm}
}
\begin{tabular}{ccc|cccc}
       &                 &         & \multicolumn{3}{c}{Operators} \\
\qquad & Energy & Config \qquad\qquad & {\tt A} & {\tt B} & {\tt C} & \qquad \\
\hline
\qquad & $-2J$           &   1    \qquad\qquad & 1  & 1 & 2 & \qquad \\
\qquad & $\phantom{-}2J$ &   2    \qquad\qquad & 1  & 0 & 0 & \qquad \\
\end{tabular}
\label{tab:matrix_sq}
\end{table}
\vspace*{3.0truecm}
\begin{table}
\caption{Probabilities for the freeze/delete operators of Fig.~\ref{ops_sq}
for the square lattice FFIM.
\vspace*{0.25truecm}
}
\begin{tabular}{ccc|cccc}
       &        &                     & \multicolumn{3}{c}{Operators} \\
\qquad & Energy & Config \qquad\qquad & {\tt A} & {\tt B} & {\tt C} & \\
\hline
\qquad & $-2J$ & 1 \qquad\qquad & $e^{-4K}$ & $p$ & $1-e^{-4K}-p$ & \\
\qquad & $\phantom{-}2J$ & 2 \qquad\qquad & 1                & 0   & 0 &   \\
\end{tabular}
\label{tab:probs_sq}
\end{table}
\begin{table}
\caption{
The average size of the maximum cluster (as a ratio of the lattice volume)
and the other clusters (given as a number of sites), and
the average number of clusters,
for different values of the free parameter $p_{sub}$ for the KBD algorithm
using the angled operations (4-7) for the square lattice FFIM at zero
temperature on a $64^2$ lattice. Also shown are the values for the
standard KBD algorithm using only operations (1-3).
\vspace*{0.25truecm}
}
\begin{tabular}{ccccc}
$p_{sub}$    	    & 0.5        & 0.8        & 1.0 	    & KBD  \\
\hline
Size of Largest Cluster  & 0.91760(2) & 0.89970(1) & 0.87362(2)  & 0.6432(5) \\
Size of Other Clusters   &   1.204    &  1.11      &  1.00       &  708      \\
Number of Clusters       & 280.22(6)  &  370.18(4) &  518.64(6)  & 2.0603(7) \\
\end{tabular}
\label{tab:max_csize}
\end{table}
\vspace*{3.0truecm}
\begin{table}
\caption{Autocorrelation times of $M^2$ for the standard KBD algorithm
applied to the square lattice FFIM at zero temperature.
\vspace*{0.25truecm}
}
\begin{tabular}{ccccc}
\qquad & $L$ &  $\tau_{int}(M^2)$  &  $\tau_{exp}(M^2)$ & \qquad \\
\hline
\qquad & ~~8 & 0.888(1)\phantom{0} & \phantom{0}0.977(9) & \qquad \\
\qquad & ~16 & 1.035(2)\phantom{0} & \phantom{0}1.85(3)\phantom{0} & \qquad \\
\qquad & ~32 & 1.265(2)\phantom{0} & \phantom{0}3.60(6)\phantom{0} & \qquad \\
\qquad & ~64 & 1.539(4)\phantom{0} & \phantom{0}5.9(1)\phantom{00} & \qquad \\
\qquad & 128 & 1.849(7)\phantom{0} & \phantom{0}9.3(2)\phantom{00} & \qquad \\
\qquad & 256 & 2.188(16)           & 13.7(5)\phantom{00}           & \qquad \\
\end{tabular}
\label{tab:tau_kbd}
\end{table}
\begin{table}
\caption{Probabilities for the freeze/delete operations of Fig.~\ref{ops_tri}
for the triangular lattice FFIM.
\vspace*{0.25truecm}
}
\begin{tabular}{ccc|ccc}
\qquad &        &        & \multicolumn{2}{c}{Operations} & \qquad \\
\qquad & Energy & Config \qquad\qquad & 1  & 2   & \qquad \\
\hline
\qquad & $-J$ & 1 \qquad\qquad & $e^{-4K}$ & $1-e^{-4K}$ & \qquad \\
\qquad & $3J$ & 2 \qquad\qquad & 1         & 0           & \qquad \\
\end{tabular}
\label{tab:probs_tri}
\end{table}
\vspace*{3.0truecm}
\begin{table}
\caption{Dynamic critical exponents for the Metropolis and plaquette
algorithms for applied to the triangular lattice FFIM at zero temperature.
\vspace*{0.25truecm}
}
\begin{tabular}{ccccc}
\qquad &  &  Metropolis  &  Plaquette & \qquad \\
\hline
\qquad & $z_{int}(\chi)$        & 0.12(5)\phantom{0}
			        & 0.04(3)\phantom{0}   & \qquad \\
\qquad & $z_{exp}(\chi)$        & 0.24(8)\phantom{0}
			        & 0.08(5)\phantom{0}   & \qquad \\
\qquad & $z_{int}(\Gamma(2))$   & 0.52(5)\phantom{0}
			        & 0.12(3)\phantom{0}   & \qquad \\
\qquad & $z_{exp}(\Gamma(2))$   & 0.70(10)
			        & 0.39(10)             & \qquad \\
\end{tabular}
\label{tab:z_tri}
\end{table}
\begin{table}
\renewcommand{\arraystretch}{0.9}
\caption{The allowance matrix for the freeze/delete operations of
Fig.~\ref{dblink_ops} for the triangular lattice FFIM at zero temperature.
\vspace*{0.25truecm}
}
\begin{tabular}{cc|ccccccccc}
        &               & \multicolumn{9}{c}{Operations} \\
Energy & Config \qquad\quad & 1 & 4 & 2 & 6 & 8 & 5 & 10 & 18 & 19    \\
\hline
$-4J$  &  1     \qquad\quad & 1 & 1 & 0 & 0 & 0 & 1 & 0  & 1  & 1     \\
$-2J$  &  2     \qquad\quad & 1 & 1 & 0 & 0 & 0 & 0 & 0  & 1  & 0     \\
$-2J$  &  3     \qquad\quad & 0 & 1 & 1 & 1 & 0 & 0 & 1  & 0  & 0     \\
$-2J$  &  4     \qquad\quad & 1 & 0 & 1 & 0 & 1 & 1 & 0  & 0  & 0     \\
$-2J$  &  5     \qquad\quad & 0 & 0 & 1 & 0 & 1 & 0 & 0  & 0  & 0     \\
$-2J$  &  6     \qquad\quad & 0 & 0 & 1 & 1 & 0 & 0 & 0  & 0  & 0     \\
\phantom{-}0\phantom{J}
       &  7     \qquad\quad & 0 & 1 & 0 & 0 & 0 & 0 & 0  & 0  & 0     \\
\phantom{-}0\phantom{J}
       &  8     \qquad\quad & 1 & 0 & 0 & 0 & 0 & 0 & 0  & 0  & 0     \\
\end{tabular}
\label{tab:dblink_matrix}
\end{table}
\vspace*{1.0truecm}
\begin{table}
\renewcommand{\arraystretch}{0.9}
\caption{Probabilities for the freeze/delete operations of
Fig.~\ref{dblink_ops} for the triangular lattice FFIM at zero temperature.
\vspace*{0.25truecm}
}
\begin{tabular}{cc|ccccccccc}
        &               & \multicolumn{9}{c}{Operations} \\
Energy & Config \qquad\quad & 1 & 4 & 2 & 6 & 8 & 5 & 10 & 18 & 19    \\
\hline
$-4J$  &  1   \qquad\quad & 0 & 0 & 0 & 0 & 0 & 0 & 0  & 0  & 1     \\
$-2J$  &  2   \qquad\quad & 0 & 0 & 0 & 0 & 0 & 0 & 0  & 1  & 0     \\
$-2J$  &  3   \qquad\quad & 0 & 0 & $p_2$ & $1\!-\!p_2$ & 0 & 0 & 0 & 0 & 0 \\
$-2J$  &  4   \qquad\quad & 0 & 0 & $p_2$ & 0 & $1\!-\!p_2$ & 0 & 0 & 0 & 0 \\
$-2J$  &  5   \qquad\quad & 0 & 0 & $p_2$ & 0 & $1\!-\!p_2$ & 0 & 0 & 0 & 0 \\
$-2J$  &  6   \qquad\quad & 0 & 0 & $p_2$ & $1\!-\!p_2$ & 0 & 0 & 0 & 0 & 0 \\
\phantom{-}0\phantom{J}
       &  7   \qquad\quad & 0 & 1 & 0 & 0 & 0 & 0 & 0  & 0  & 0     \\
\phantom{-}0\phantom{J}
       &  8   \qquad\quad & 1 & 0 & 0 & 0 & 0 & 0 & 0  & 0  & 0     \\
\end{tabular}
\label{tab:dblink_ops}
\end{table}
%
%

%
%
% FIGURES
%
% ********* section square **************
%
\figure{
Freeze/delete operations for the square lattice FFIM.
The bold lines indicate frozen bonds.
\label{ops_sq}}
\figure{Log-log plot of the autocorrelation times of $M^2$ for
the standard KBD algorithm for the square lattice FFIM at zero temperature.
\label{tau_kbd}}
\figure{Log-log plot of $\tau_{int}$ of $M^2$ for the square lattice FFIM
at zero temperature, using the Metropolis algorithm,
the standard KBD algorithm, the new improved KBD algorithm
where a cluster is flipped at every iteration, and the KBD algorithm using just
angled operations with three different values of the free parameter $p_{sub}$.
\label{tau_all}}
\figure{Autocorrelation function for the improved KBD algorithm
for the FFIM on a $64^2$ lattice at zero temperature.
\label{new_kbd}}
\figure{Autocorrelation functions of $M^2$ for the square lattice
FFIM at zero temperature on a $64^2$ lattice for the KBD algorithm
with and without Metropolis sweeps, with asymptotic fits to an
exponential and a logarithm respectively.
\label{rho_ergodic}}
%
%
% ********* section triangle **************

%
\figure{Freeze/delete operations for the triangular lattice FFIM.
The bold lines indicate frozen bonds.
\label{ops_tri}}
\figure{Autocorrelation times for $\chi$ at $T\!=\!0$ for the
triangular lattice FFIM using the Metropolis and plaquette algorithms.
\label{tau_chi}}
\figure{Autocorrelation times for $\Gamma(2)$ at $T\!=\!0$ for the
triangular lattice FFIM using the Metropolis and plaquette algorithms.
\label{tau_gamma}}
\figure{Extensions of basic triangular plaquette:
(a) double plaquette, (b) double plaquette plus extra link.
\label{dbl_tri}}
\figure{Possible ground state configurations of the tile shown in
Fig.~\ref{dbl_tri}(b). Here solid lines denote satisfied bonds.
\label{dblink_configs}}
\figure{Possible energy conserving operations for the tile shown in
Fig.~\ref{dbl_tri}(b).
The bold lines indicate frozen bonds.
\label{dblink_ops}}

\end{document}